\documentclass{aastex}
\usepackage{myemulateapj5}
\begin{document}
\vspace*{\fill}
\title{\emph{HST} ABSOLUTE SPECTROPHOTOMETRY OF VEGA FROM THE FAR-UV TO THE IR}
\author{R.~C.\ Bohlin  \&  R.~L.\ Gilliland} 
\affil{Space Telescope Science Institute\\
3700 San Martin Drive\\
Baltimore, MD 21218\\
bohlin@stsci.edu, gillil@stsci.edu}
\vspace{1in}
\centerline{\titlelarge To be published in the 4 June 2004 issue of the \emph{Astronomical Journal}}
\received{2 February 2004}
\accepted{19 February 2004}
\sluginfo
\vfill

\begin{abstract} The Space Telescope Imaging Spectrograph has measured the
absolute flux for  Vega from 0.17--1.01~$\mu$m on the \emph{HST} White Dwarf
flux scale. These data are saturated by up to a factor of 80 overexposure but
retain linearity to a  precision of 0.2\%, because the charge bleeds along the
columns and is recovered  during readout of the CCD. The S/N per pixel exceeds
1000, and the resolution $R$ is about 500. A $V$~magnitude of 0.026$\pm$0.008
is established  for Vega; and the absolute flux level is
$3.46\times10^{-9}$~erg cm$^{-2}$ s$^{-1}$ at 5556~\AA. In the regions of
Balmer and Paschen lines, the STIS equivalent widths differ from the pioneering
work of Hayes in 1985 but do agree with predictions of a Kurucz model
atmosphere, so that the STIS flux distribution is preferred to that of Hayes.
Over the full wavelength range, the model atmosphere calculation shows
excellent agreement with the STIS flux distribution and is used to extrapolate
predicted fluxes into the IR region. However, the IR fluxes are 2\% low with
respect to the standard Vega model of Cohen. \emph{IUE} data provide the
extension of the measured STIS flux distribution from 0.17 down to 0.12~$\mu$m.
The STIS relative flux calibration is based on model atmosphere calculations of
pure hydrogen WDs, while the Hayes flux calibration is based on the physics of
laboratory lamps and black body ovens. The agreement to 1\% of these two
independent methods for determining the relative stellar flux distributions
suggests that both methods may be correct from 0.5--0.8~$\mu$m and adds
confidence to claims that the fluxes relative to 5500~\AA\ are determined to
better than 4\% by the pure hydrogen WD models from 0.12 to 3~$\mu$m. 
\end{abstract}  

\keywords{stars: individual (Vega) ---
techniques: spectroscopic ---  stars: fundamental parameters (absolute flux)}

\section{INTRODUCTION}

The most commonly used flux calibration for the fundamental standard Vega
($\alpha$~Lyr, HD~172167, HR~7001) is the compilation of Hayes (1985), while
Megessier (1995) suggests an increase of 0.6\% in the Hayes value of
$3.44\times10^{-9}$ erg cm$^{-2}$ s$^{-1}$ \AA$^{-1}$ at 5556~\AA. The Hayes
flux for Vega is traceable to fundamental standard lamps pedigreed by NBS (now
NIST), which is charged with maintaining fundamental standards of physical
units. The Hayes flux compilation is based on direct comparisons of the star to
calibrated lamps and black body ovens. Most ground based estimates of absolute
stellar flux are traceable to the flux of Vega.

With the advent of extensive space based observations in the far-UV, a need for
standard flux candles arose at wavelengths below the atmospheric cutoff. A few
rocket experiments established some UV standard stars with $\sim$10\% precision
(e.g.\  Strongylis \& Bohlin 1979). To establish standards with better
accuracy, a technique based on model atmospheres for simple, pure hydrogen WDs
was suggested by D.~Finley and J.~Holberg for calibrating the \emph{IUE}
satellite spectrophotometry, (cf.\ Fig.~8 in Bohlin, et~al.\ 1990). The hot WD
standard stars G191B2B, GD~153, and GD~71 at $V=12$--13~mag were established by
computing LTE model atmosphere flux distributions for pure hydrogen
atmospheres, which were then normalized to precisely measured $V$~magnitudes
from A.~Landolt (see Bohlin 2000). EUV observations below the Lyman limit
demonstrated that any effects of interstellar extinction in these three stars
is less than 1\% longward of the 912~\AA\ Lyman limit. These three WD standards
are now based on NLTE models by Ivan Hubeny (Bohlin 2003) and are
complemented by three solar analog stars for \emph{HST} calibrations extending
into the IR (Bohlin, Dickinson, \& Calzetti 2001). 

The Vega observations with STIS are discussed in Section~2. Section~3
establishes the best estimate of the V band magnitude of Vega, while Section~4
presents the first direct comparison between the WD and the standard lamp
methods for establishing absolute flux. A third method for defining the flux of
Vega at all wavelengths is based on modeling of its stellar atmosphere, as
discussed in Section~5. Model atmosphere codes are now sophisticated enough to
warrant detailed comparisons between observation and theory. Because of the
excellent agreement of a model (Kurucz 2003, Castelli and
Kurucz 1994) with the STIS observations up to
1~$\mu$m, the model should provide a good estimate of the flux beyond 1~$\mu$m,
as discussed in Section~6.

\section{CALIBRATION OF THE SATURATED OBSERVATIONS}

STIS observations of Vega in the three CCD low dispersion modes G230LB, G430L, 
and G750L (Kim Quijano 2003) produce heavy saturation of the full well depth.
However at Gain~=~4, the A-to-D amplifier does not saturate; and the excess
charge just bleeds into adjacent pixels along the columns, which are
perpendicular to the dispersion axis. Gilliland, Goudfrooij, and Kimble (1999,
GGK) have demonstrated that saturated data on the STIS CCD are linear in total
charge vs.\ stellar flux, as long as the extraction region on the CCD image is
large enough to include all the charge. In particular, GGK demonstrated
linearity to 0.1\% accuracy using 50$\times$ overexposed images of a star in
M67 to compare with unsaturated exposures of the same star. In the G430M medium
dispersion spectral mode, 0.8s observations of $\alpha$~Cen~A saturated by
about a factor of 4; and GGK showed that the ratio to a mean G430M spectrum at
0.1s is linear to $\ll1$\%. Furthermore, GGK demonstrated that the short 0.3s
STIS exposure times are stable and accurate to 0.2\%, i.e.~0.0006s.

\subsection{Corrections for the Tall Extractions of the Saturated Data}

For Vega, the extraction heights required to include all the charge in the
saturated images are 84, 54, and 40 pixels tall for G230LB, G430L, and G750L,
respectively, for single integrations of 6s for G230LB and 0.3s for G430L and
G750L. For these saturated images, the signal level is so high that any signal
loss due to charge transfer efficiency (CTE) effects (Bohlin \& Goudfrooij
2003) is $<$0.1\%.

Because of scattered light from the STIS gratings, a tall extraction height
contains more signal than the standard 7~pixel extraction, to which the
absolute flux calibrations apply (Bohlin 1998). The amount of excess signal in
the tall extractions must be derived from unsaturated images. Observations of
the star AGK+81$^\circ$266 that is used to monitor the change in STIS
instrumental sensitivity are used to derive the correction for the heights of
54 and 40 pixels, respectively for G430L with 43~observations and G750L with 
37~observations. For G230LB, the three unsaturated G230LB observations of Vega
itself are used to correct the 84~pixel extractions of the saturated
observations. All three unsaturated Vega G230LB spectra at the detector center
give the same correction from a 7 to an 84~pixel high extraction. This
correction drops from 1.293 at 1700~\AA\ to 1.150 at 3000~\AA\ with a maximum
difference among the three observations of 0.2\% in the 1.293 correction factor
at 1700~\AA. 

The uncertainties in the tall slit corrections for G430L and for G750L below
8450~\AA, where saturation ends (see 2.3 below), can be characterized by the
rms scatter among the many AGK+81$^\circ$266 determinations. The ratios of the 54 and
40 pixel high extractions to the standard 7~pixel extraction have an added
uncertainty of 0.2--0.6\% measured by the rms of these ratios for AGK+81$^\circ$266
but do not show trends with time that might be caused by the increasing CTE
losses. The CTE correction for the 7~pixel extractions grows from zero in 1997
to a maximum of 3\% in the relevant 3000--8450~\AA\ range for the
AGK+81$^\circ$266 data obtained near the time of the Vega observations; but the
limit to any error caused by our assumption of a matching CTE correction in the
54 or 40~pixel tall extractions is included in the 0.2--0.6\% rms values. For
example in the range covered by the $V$~filter (see Section~3), the rms
uncertainty in the tall slit correction is 0.3\%. The rms uncertainties of
$<$0.2\% for G230LB and 0.2--0.6\% over the 3000--8450~\AA\ range for the tall
slit correction ratios should be included in the total systematic error budget
as a function of wavelength.

\subsection{Verification of Linearity for Saturated Data}

\subsubsection{Shutter Timing}

Linearity for the Vega data can be verified directly, because 0.3s observations
of Vega in the G230LB mode are unsaturated below 2900~\AA, i.e.\ over the first
900 pixels of the 1024 pixel spectrum. Exposures of 6s were obtained in order
to re-verify linearity up to a saturation factor of 20. The 6s observations
with G230LB are more saturated at the longest wavelengths than at any
wavelength in the G430L or G750L spectra. Table~1 is the journal of the Vega
observations, while Figure~1 demonstrates both the repeatability of G230LB
observations and, also, the linearity beyond saturation. The individual frame
exposure times are the exptime per cr-split frame from Table~1, i.e.\ 0.3s for
all observations, except for the saturated G230LB observations of 6s. One
spectrum in the 52X2E1 aperture was placed 130~pixels from the top of the
detector, where the CTE correction is 4$\times$ lower than the central
correction, which exceeds 1\% only below 2000~\AA. However, the flat field
correction for the E1~position relative to centered spectra is currently based
on only 13~observations obtained in 2000--2002 and is uncertain by $\sim$0.5\%.
The E1~exposure will not be included in the analysis below. Figure~1 shows the
ratio of the five centered individual G230LB observations to their average. The
two 6s exposures are low by 0.9998$\pm$0.0005 and the three centered and
unsaturated ratios average 1.0027$\pm$0.0005. The discrepancy between the flux
for the two exposure times is 0.29\%$\pm$0.07\%. One simple explanation is that
the nominal 0.3s exposures are really 0.3009s$\pm$0.0002s, in agreement with
GGK who find that the 0.3s shutter timing is accurate to $\sim$0.0006s.

\subsubsection{Saturation and Charge Transfer Efficiency Corrections}

Other possibilities for the 0.3\% flux difference between the 0.3 and 6s
exposures are non-linearities, either in the saturated data or in the CTE
correction of the unsaturated data. GGK demonstrate linearity to better than
0.1\% for exposures of point sources overexposed by 50$\times$ and for spectra
of $\alpha$~Cen~A at 4$\times$ over saturation (see above). The loss of signal
due to charge transfer efficiency (CTE) is $<$0.1\% for the saturated spectra,
while the amount of the CTE correction for the unsaturated 0.3s G230LB data
ranges from $\sim$1.5\% at the shorter wavelengths to $\sim$0.5\% at the
longest wavelengths. Bohlin and  Goudfrooij (2003) suggest their CTE correction
formula is accurate to 10\% for the signal level of the Vega data, which
corresponds to $<$0.2\% for the CTE correction of $<$2\% for the 0.3s fluxes.
Because these uncertainties are comparable to the measured timing error of the
0.3s exposures, the nominal 0.3000s value is retained. The uncertainty of 0.3\%
in the 0.3s exposure times will contribute to the uncertainty in the final
determination of the measured STIS $V$~mag that will be compared with
$V=0.035$, which was previously used to set the overall level of the absolute
STIS flux calibration on the WD scale (Bohlin 2000 and Section~3 below.)

\subsection{Additional Details}

The 84 pixel extractions have more contamination from the out-of-band scattered
light that fills in the line cores slightly and produces the $\sim$1\% dips in
Figure~1 at the locations of the deeper lines in the 7~pixel extractions from
the 0.3s images. At the strong 2800~\AA\ \ion{Mg}{2} line, the differential
filling reaches $\sim$2\%. To minimize the out-of-band stray light, a standard
7~pixel extraction is used longward of 8450~\AA, where there is no saturation
in the 0.3s exposures with G750L. 

After extracting the spectra from the images, adjusting the flux to a standard
7-pixel-high aperture, correcting for sensitivity changes with time Bohlin
(2003), and applying the CTE correction (Bohlin \& Goudfrooij 2003),
corrections to the wavelengths are made for sub-pixel wavelength errors that
are obvious in the high S/N saturated spectra. These shifts range up to
0.5~pixel and are found by cross-correlation of the absorption lines with a
model flux distribution.

\section{THE \emph{V} BAND MAGNITUDE OF VEGA}

The STIS flux calibration is based on standard star reference fluxes that are
pure hydrogen WD models normalized relative to the average flux
of Vega over the $V$~band (Colina and Bohlin 1994, Bohlin 2000). The absolute
flux of these fundamental standard stars G191B2B, GD~71, and GD~153, depends
directly on the difference between the $V$~magnitude of Vega and $V$ for each
WD star. The one sigma uncertainties in the Landolt $V$~magnitudes of the WD
stars are in the range 0.001--.004 mag (Landolt 1992 and private communication).
The measured STIS flux for Vega relative to the WD standard star fluxes
determines the instrumental $V$~magnitude of Vega with respect to the precision
Landolt $V$~magnitudes of the WDs. The Landolt transformations from his
instrumental scale to Johnson $V$~magnitudes is summarized in Colina and Bohlin
(1994). There are three choices for the definition of the Landolt $V$~bandpass:
the Kitt Peak filter+detector (Landolt, private communication), the CTIO
filter+detector (Landolt 1992), and the latter as modified by Cohen et~al. \
(2003b) to include an estimate of the atmospheric transparency above the CTIO
observatory. None of the estimated bandpass functions include the telescope
optics; but that omission is expected to cause $<$0.001 mag errors (Colina and
Bohlin 1994). The three bandpass functions yield slightly different magnitudes
for Vega in the range of $V=0.021$ to 0.031; but $V=0.026$ for the Cohen et~al.\
(2003b) bandpass
with the atmosphere included should be the most accurate. Conservatively,
0.005~mag is assigned to the uncertainty that still remains in the
Landolt-Cohen CTIO bandpass function. The small scatter of 0.002 among the
separate determinations from the three WDs is included in the 0.005~mag
bandpass error estimate.

Colina and Bohlin (1994) previously adopted $V=0.035$ for Vega. Megessier (1995)
and Bessell, Castelli, and Plez (1998) adopt $V=0.03$, while Johnson et~al.\
(1966) also quote 0.03 in their Table~2 of the best mean values with a probable
uncertainty of 0.015~mag. Because the STIS determination is more accurate and
differs from the ground based value by less than its $\sigma=0.015$~mag,
$V=0.026\pm0.008$ is adopted for Vega. The rms uncertainty of 0.008 is the
combination of the 0.005~mag from the bandpass and 0.003~mag uncertainties from
each of the shutter timing, the tall slit correction, the Landolt WD
magnitudes, and the CTE correction to the WD standard stars. The constant
offset of 0.006~mag between Johnson~$V$ and color corrected Landolt~$V$ is not
usually applied (Colina and Bohlin 1994). However, the 0.006 has been added for
this direct comparison to Johnson~$V$ values, while 0.020 is used to compare with
the Landolt~$V$ magnitudes of the WD standard stars. Because the STIS
observations determine the flux of Vega relative to the fundamental 12--13~mag
WD standards, adopting $V=0.026$ means that the STIS low dispersion
spectroscopic modes on the CCD detector are perfectly linear over this
$\sim$10$^{^5}$ range in stellar flux.

\section{COMPARISON OF STIS FLUXES WITH HAYES}

Megessier (1995) claims that the average monochromatic Hayes flux of
$3.44\times10^{-9}$~erg cm$^{-2}$ s$^{-1}$ \AA$^{-1}$ at 5556~\AA\ includes one
value from a faulty tungsten strip lamp at Palomar. Eliminating the Palomar
value of $3.36\times10^{-9}$~erg cm$^{-2}$ s$^{-1}$ \AA$^{-1}$, Megessier finds
that the flux at 5556~\AA\ is $3.46\times10^{-9}$~erg cm$^{-2}$ s$^{-1}$
\AA$^{-1}\pm0.7$\% from a weighted average of three remaining independent
determinations. The absolute scaling of the STIS sensitivity functions is set
by a recalibration for $V=0.026$ and the corresponding integration over the
model as normalized to $3.46\times10^{-9}$~erg cm$^{-2}$ s$^{-1}$ \AA$^{-1}$ at
5556~\AA. Even though the monochromatic model flux is increased by only 0.6\%
at 5556~\AA\ relative to Hayes (1985), the slightly different flux distributions
over the broad band V filter causes a 0.9\% higher average for the model
compared to the Hayes flux distribution. The resulting STIS flux for Vega
produces a ratio of STIS divided by the original Hayes (1985) fluxes shown by
the solid line in Figure~2. Before dividing by the Hayes flux, the STIS values
are binned over the Hayes 25~\AA\  bandpass, the ratio is then smoothed twice
with a 5~point boxcar and plotted at the 25~\AA\ sample spacing of the Hayes
compilation. 

More important than the question of absolute flux level is the slope
(dF/d$\lambda$) of the independently determined WD and Hayes flux
distributions. Over the 5000--8500~\AA\ range, the two flux distributions agree
to $\sim$0.5\% in slope. The H$\alpha$ lines of Hayes and STIS also match; but
the higher Balmer lines below 5000~\AA\ and the Paschen lines above 8500~\AA\
are significantly different in the two spectra. In order to understand the
nature of these differences, the two measured fluxes are compared with the
Kurucz (2003) $R=500$ model in Figures~3--4 on a vacuum wavelength scale. The
resolving power $R$ of the three STIS CCD low dispersion modes ranges from 500
to 1100 for narrow slits; however, the broader PSF in the 2~arcsec wide slit
and the tall extraction heights for the saturated data cause some added
degradation. The line depths of the hydrogen lines in the STIS spectra are best
matched by the Kurucz model at $R=500$. Figure~3 illustrates the discrepant
Hayes H$\beta$ line profile in comparison to the model and to STIS. The
equivalent width of the Hayes H$\beta$ line is too low. The rest of higher
Balmer lines show similar discrepancies, which cause the three big dips in
Figure~2 between 3900 and 5000~\AA, where STIS measures stronger features.
Perhaps, the Hayes data at the shorter wavelengths suffer from stray light that
fills in the line profiles. Figure~4 illustrates a different situation for the
Paschen lines. At these long wavelengths, STIS CCD observations suffer from a
severe fringing pattern that must be removed with a contemporaneous flat from a
tungsten lamp. The de-fringed STIS spectrum suffers from residual artifacts at
the $\sim$2\% level, so that all the differences between the STIS data and the
model can be ascribed to fringe residuals. Several of the Hayes lines are too
strong by up to a factor of~2; and the centroids of several of his spectral
features lie at the wrong wavelengths. In one case, the 8677.4~\AA\ (8675~\AA\
air wavelength) Hayes point is a peak, rather than the minimum that should
correspond to the Paschen line just shortward, at 8667.4~\AA. These apparent
major errors in the Hayes spectrophotometry cause the spurious structure in the
ratio of STIS/Hayes in Figure~2. Hayes (1985) used the 25A step size from the 
one available data set with a continuous scan but also included other data sets
with resolutions of 10 to 100~\AA\ in his final weighted average of absolute
fluxes. He warns of the low accuracy of his energy distribution ``...near strong
lines and in the Balmer and Paschen confluences...'' Some additional problems
with the Hayes spectrophotometry at the longer wavelengths of Figure~4 are
probably caused by telluric water vapor lines, while the Balmer line region may
be contaminated by ozone and $O_\mathrm{4}$ absorption features in the Earth's
upper atmosphere (Kurucz 1995). Our final flux distribution (see the next
section for details) is shown in
Figure~5 and is entirely independent of the Hayes compilation.

\section{COMPARISON OF THE STIS OBSERVATIONS WITH A VEGA MODEL ATMOSPHERE CALCULATION}

The excellent agreement of the STIS spectrum with the Kurucz (2003)
calculations motivates a more detailed investigation with the goal of
establishing typical uncertainties that might apply to the model atmosphere
fluxes in the unobserved IR region. The Balmer line profiles are correct in the
Kurucz models, because the effective temperature (9550~K) and gravity ($\log 
g=3.95$) are determined by fitting theoretical line profiles to high resolution
observations (Castelli \& Kurucz 1994). The metallicity is [M/H]~$=-0.5$, and
a detailed discussion of the elemental abundances is in Castelli and Kurucz
(1994). The main difference between the Kurucz 
(2003) and the Castelli and Kurucz (1994) models is the correction of a spurious
dip in the flux distribution in the 3670--3720~\AA\ region.

\subsection{6600--8500 \AA}

Starting the detailed comparison at the long wavelengths, only one spectral
feature, the \ion{O}{1} multiplet at 7775~\AA\ (7773~\AA\ air), is deeper
than 1\% in the range from H$\alpha$ to the 8500~\AA\ cutoff of Figure~4. The
only deviation $>$1\%  between the model and the STIS flux is from
7800--8200~\AA, where STIS is systematically $\sim$1\% high. Again, this small
difference is probably an artifact of the de-fringing process.

\subsection{Scaling to the Continuum Level}

Below H$\alpha$, there are no fringes in the STIS spectra; and Figures~6--8
show the comparison between the data and the model after normalizing to a
continuum. These figures show the fluxes divided by the theoretical smooth
continuum, conveniently provided by Kurucz as a third column in his model
files. Both STIS and the model fluxes are divided by the same continuum
values. The continuum level has no physical significance and only provides a
scaling that allows the spectral features to be displayed on a common expanded
scale. Therefore, from 3640--3980~\AA, a lower continuum level is adopted in
order to bring the features in this range closer to unity after normalization.
Figure~6  presents the regime from H$\delta$ to H$\alpha$; Figure~7 covers the
Balmer continuum and higher Balmer lines; and Figure~8 compares the
observations with the model in the region with strong metal line blanketing
from \ion{Mg}{2} (2800~\AA) down to the STIS G230LB cutoff. 

\subsection{4100--6600  \AA}

Figure 6 shows the region of best agreement, where model and observation agree
to $\sim$1\%. The two 1--2\% differences in the centers of weak lines could be
due to small differences of the actual STIS resolution from the $R=500$ model. 

\subsection{2800--4100 \AA}

Figure 7 shows the most troubling divergence of the model from the STIS flux
distribution. The model has higher fluxes by $\sim$3\% from $\sim$3100~\AA\ to
3670~\AA. This difference gradually diminishes toward shorter wavelengths and
disappears around 2800~\AA, while a similarly larger model flux appears in
the five or six well resolved regions between the Balmer lines shortward of
H$\delta$. In the problematic region of difficult atomic physics (Castelli and
Kurucz, see above), the model still dips for 20--30~\AA\ near 3700~\AA\ to a
few percent lower than the smoother STIS transition from lines to continuum.
Because the STIS flux calibration is based on similar calculations of the shape
of the Balmer jump region, the source of the difference must arise from errors
in the WD models or in the Kurucz Vega model. The Balmer jump in the hot WDs is
6--7 times weaker; and line strengths are correspondingly less, so that the
hotter WD fluxes should have a lower fractional error across the transition
region from Balmer lines to continuum. In other words, any opacity errors in
the Balmer continuum or in the overlapping wings of the lines are magnified
in the Vega model relative to a 30,000--60,000K WD model atmosphere.

\subsection{1680--2800 \AA}

Figure 8 covers the region of heavy metal line blanketing and shows excellent
agreement of the wavelengths of the spectral bumps and dips, implying that the
model does include the proper set of spectral lines. Several regions agree in
flux to 1\%, especially those around the 0.9 level, i.e.\ 10\% blanketing.
However, as the blanketing increases, the amount of increase is too large in
the model. For example, the model is too low by an average of 6\% in the
1800--1900~\AA\ region, where the STIS flux drops as low as 0.82 compared to
0.74  for the model. The source of these differences are difficult to ascribe
to STIS, because its flux calibration is determined by the WD standard stars,
which are perfectly smooth with no lines in this region. Broadband residuals in
the STIS calibration are $<$0.5\% for all three standard stars between
Ly-$\alpha$ and 9600~\AA, while narrow band ($\sim$25~\AA) residuals are all
$<$1\%. Because the differences in Figure~8 increase with the amount of
blanketing, a more likely cause seems to be some systematic overestimate of the
line strengths in the model. There is no evidence for interstellar reddening,
which would reduce the observed flux the most in the 2200~\AA\ region for the
standard galactic reddening curve of Seaton (1979).

\subsection{Below the 1680 \AA\ Cutoff of STIS}

In the far-UV, the comparison of the model with the \emph{IUE} spectrum follows
the above pattern, where the model is often lower but rarely higher than the
measured flux distribution. An outstanding example is from 1600--1650~\AA,
where the \emph{IUE} is fairly constant at 0.80 to 0.85 relative to the
continuum. However, the model has a big dip down to 0.64 and averages
$\sim$0.77, i.e.\ the model is $\sim$7\%  low averaged over 1600--1650~\AA. The
possibility that the UV metal lines are too strong in the model is again
supported by a comparison of the strong 1670.8~\AA\  \ion{Al}{2} line in the
model vs.\ an \emph{IUE} high dispersion spectrum. The full width at half the
continuum level of the \emph{IUE} line is 0.81~\AA, while the measured width in
the model is 0.88~\AA, even when comparing the full resolution $R=500,000$
Kurucz spectrum to the $R=\sim$12,000 \emph{IUE} line profile.

\subsection{The Recommended Flux Distribution}

\emph{IUE} spectrophotometry on the \emph{HST} WD scale (Bohlin 1996) completes the
observed flux distribution of Vega below the STIS CCD cutoff at 1675~\AA\ after
multiplying by 1.04 to normalize to the STIS fluxes over the 1700--1900~\AA\
range. Longward of $\sim$4200~\AA, the few isolated differences between the
STIS spectrum and the model are at the 1--2\% level and can be attributed to
instrumental effects. In the continuum, the STIS data are 0.4\% higher at 5556~
\AA\ and only 0.1\% higher in the $V$~band. Therefore, in view of this excellent
agreement of the model with STIS, the $R=500$ model with effective temperature
9550~K and $\log  g=3.95$ is adopted as the standard star flux distribution
longward of 4200~\AA. The STIS PSF is expected to have a Gaussian core
corresponding to $R \sim500$ with broad wings extending to the 2~arcsec width
of the 52$\times$2 entrance slit. The $R=500$ model that matches the STIS resolution is
shown in Figure~5 and comprises the recommended flux distribution. However,
higher resolution model spectra are available from Kurucz (2003) and provide
equally valid model flux distributions after normalization to
$3.46\times10^{-9}$~erg cm$^{-2}$ s$^{-1}$ \AA$^{-1}$ at 5556~\AA. From 1675 to
4200~\AA, the final Vega absolute flux distribution is defined by the new STIS
observations as calibrated on the \emph{HST} pure hydrogen WD scale. The only
dependencies of this \emph{HST} flux scale on ground based observations are the
adopted monochromatic flux for Vega at 5556A from Megessier (1995) and the
Landolt V~magnitudes of the three pure hydrogen primary standard WDs, as
reviewed in the Introduction but refined using the new V~magnitude and model
flux distribution defined for Vega in this work. The URL for accessing the new
Vega flux distribution is\hfill\break
http://www.stsci.edu/instruments/observatory/ cdbs/calspec. The
file name is\hfill\break alpha\_lyr\_stis\_002.fits. Included in the file are the estimated
systematic uncertainty of 1\% and the statistical uncertainties, which
determine a S/N per pixel that ranges from 1300 at 1700~\AA\ to 6700 at
3000~\AA.

\section{THE INFRARED SPECTRUM OF VEGA}

Our infrared spectrum of Vega is provided by the $R=500$ Kurucz model, after
normalizing the model to the $3.46\times10^{-9}$ flux of Megessier at 5556~\AA.
Cohen  et~al.\ (1992) used 1991-vintage Kurucz models to establish Sirius and
Vega as primary IR irradiance standards, which are the basis for a heroic
effort to standardize IR fluxes and photometry that is now up to Paper XIV
(Cohen et~al.\  2003a). While our Vega flux distribution is 0.6\% higher at
5556~\AA, the cooler Cohen temperature of 9400~K makes the broadband ratio of
the 9550/9400~K models about 2\% lower from 1.2 to 35~$\mu$m. Our normalization
of the hotter Kurucz model implies an angular diameter of 3.273~mas, in
comparison with the Cohen et~al.\ (1992) value of 3.335~mas and the measured
value of 3.24$\pm$0.07~mas (Hanbury Brown et~al.\ 1974). However, a 2\% lower
IR flux for Vega causes worse agreement with the few existing absolute IR flux
measurements as summarized by
Cohen et~al.\ (1992), who adopted an uncertainty in the IR of 1.5\% from the
Hayes (1985) standard error at 5556~\AA. Perhaps, 2\% is a more realistic
minimum uncertainty for Vega in the IR. Because the more recent Kurucz LTE
models have realistic line profiles, narrow band comparisons of the 1991 vs.\
2003 Kurucz models can differ by larger factors, if an absorption line is an
important contributor to the average flux over the bandpass.

Future work on Vega model atmosphere calculations might solve the discrepancies
below 4100~\AA\ (see Section~5) and could also result in a cooler
$T_\mathrm{eff}$ than used by Kurucz (2003). The 2\% IR uncertainty for the
Vega modeling effort is not applicable to the pure hydrogen WD calibrations,
where lack of metal line absorption and higher $T_\mathrm{eff}$ result in lower
dispersion between the possible WD IR flux distributions below $\sim$2~$\mu$m.
Above 2~$\mu$m, differences between LTE and NLTE model atmospheres for the
hottest WD~G191B2B exceed 2\% (Bohlin 2003). Because NLTE models for the hot
WDs are brighter in the IR compared to the visible, a NLTE model for Vega might
improve the agreement with the IR absolute flux determinations.

The use of a model to represent the IR flux from Vega does not produce an ideal
standard star. Cohen et~al.\ (1992) utilize Sirius as their only primary
standard beyond 17~$\mu$m because of Vega's disk of cold dust. Furthermore,
Gulliver, Hill, \& Adelman \ (1994) point out that Vega is a pole-on rapid
rotator. Kurucz tuned his Vega model to fit the average visible spectrum.
However, the hot poles emit more in the UV; the cooler equator has more flux in
the IR; and the averages in the UV and IR may be appropriate to different
models. A composite average of Kurucz models for $T_\mathrm{eff}=9300$~K at the
equator to 9620~K at the pole has been constructed by Hill, Gulliver, and
Adelman (1996) and is currently being updated by Gulliver (private
communication). Hopefully, the composite model will be brighter in the UV and
IR relative to 5556~\AA\ and will fit the absolute STIS spectrophotometry and
the absolute IR flux measurements better, while maintaining the same 5556~\AA\
surface brightness as the 9550K model that produces the excellent agreement
with the Hanbury Brown angular diameter.

Sirius is a slow rotator and easier to model, because one model should fit all
wavelengths. STIS observations of Sirius should be made to test whether the 
Kurucz (2003) model for Sirius and its observed flux agree better than for the
Vega case. In addition, more high precision IR absolute flux measurements
should be made relative to pedigreed laboratory light sources.

\section{SUMMARY}

As a result of the STIS observations of Vega, a V mag of 0.026 is established
for the improved bandpass function of Cohen et~al. \ (2003b). The Kurucz \
(2003) model of Vega agrees so well with the STIS observations that the model
itself defines the standard star fluxes longward of 4200 \AA. These
improvements, combined with the more realistic bandpass function that is used
to translate the V magnitudes of the WD standards into relative fluxes, have
resulted in WD fluxes that are uniformly 0.5\% fainter than found by Bohlin \
(2000). The WD standard star model flux distributions must be updated in the
CALSPEC archive, along with the new Vega spectrum. Because these files have not
been updated since 00Apr4, the changes quantified by Bohlin (2003) as a
function of wavelength for the switch from LTE to NLTE are also included in the
new *mod*.fits files. The updates for the CALSPEC secondary flux standards are
being prepared and will reflect the above changes plus a number of new STIS
observations, better corrections for the changes with time and temperature, and
improved CTE corrections for the CCD data. In addition, a suite of NICMOS grism
calibration observations are being obtained to resolve the IR differences
between the pure hydrogen WDs and the solar analogs and to extend the
\emph{HST} spectrophotometric coverage to 2.5~$\mu$m. Unfortunately, Vega is
too bright for NICMOS grism observations.

\acknowledgements Robert Kurucz and Martin Cohen provided extensive comments that
have been incorporated in this paper.
Primary support for this work was provided by NASA through 
the Space Telescope Science Institute,
which is operated by AURA, Inc., under NASA contract NAS5-26555. Additional
support came from DOE through contract number C3691 from the
University of California/Lawrence Berkeley National Laboratory.

\begin{figure}
\centering 
\includegraphics[height=6in]{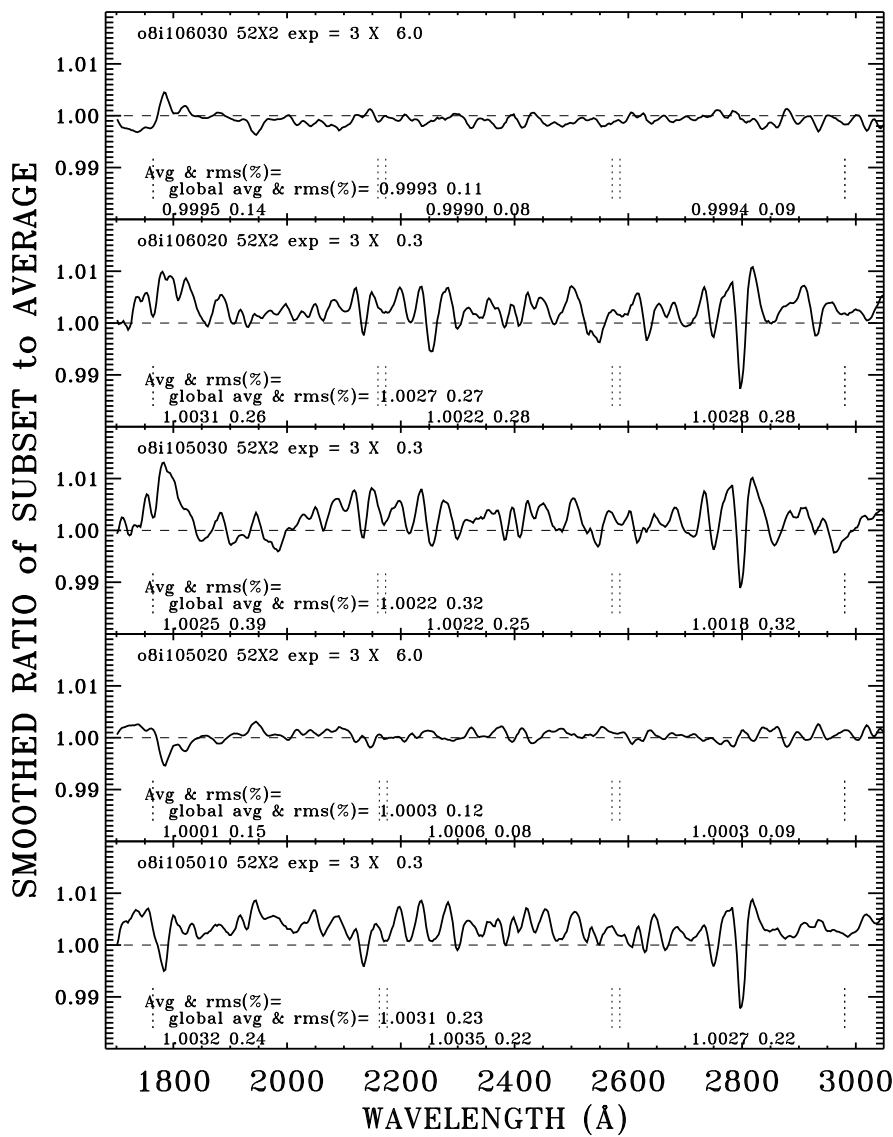}
\caption{
Ratios of each of the five centered observations of
Vega in the G230LB mode to the average of these five spectra.  The total
exposure time of the CR-split=3 observations are written on the plots. The
average spectrum is dominated by the two heavily saturated observations of 18s,
while the shorter observations are not saturated below 2900~\AA. The three
short observations are all 0.2--0.3\% brighter than the
average and show a flat ratio from 1900 to 3000~\AA, corresponding to a range
in detected photoelectrons of 7$\times$. The strongest 
features in the short exposure
ratios are absorption lines that have slightly more contamination by
out-of-band light in the tall extraction height of 84 pixels required to catch
all the saturation. For example, \ion{Mg}{2} at 2800~\AA\ has a deeper line by
$\sim$2\% in the 7~pixel extraction of the 0.9s observation relative to the
84~pixel extraction of the 18s data. The global average and rms of the
residuals are written on each panel, along with three mean and rms values for
the three separate regions delineated by the vertical dashed lines.}
\end{figure}

\begin{figure}
\centering
\includegraphics[width=.5\textwidth,angle=90]{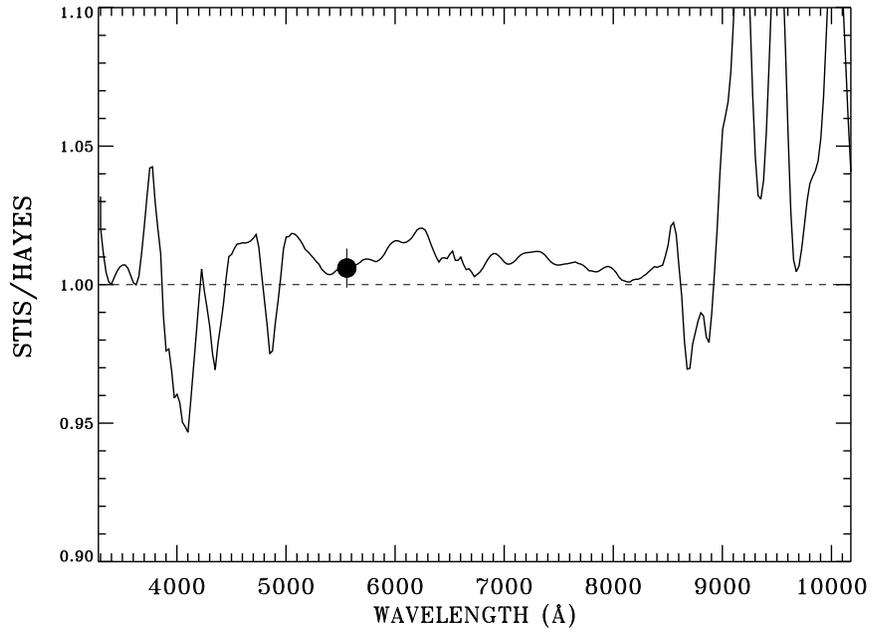}
\caption{Ratio of the final STIS fluxes for Vega to those of
Hayes (1985). The revised monochromatic flux of Megessier (1995) at 5556~\AA\
is shown as the filled circle.}
\end{figure}

\begin{figure}
\centering
\includegraphics[width=.5\textwidth,angle=90]{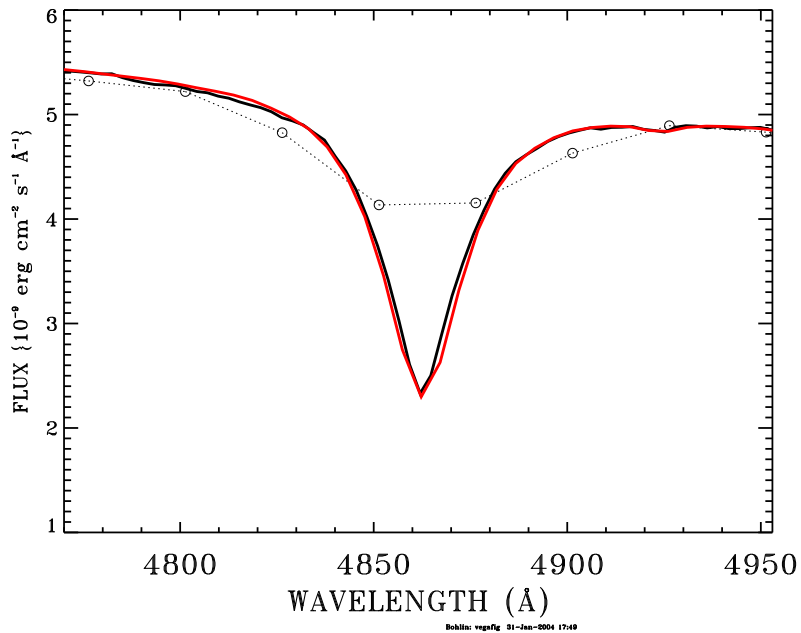}
\caption{The H$\beta$ spectral region. Solid line: final STIS flux, 
red line: Kurucz (2003) $R=500$ model scaled to 
$3.46\times10^{-9}$~erg cm$^{-2}$ s$^{-1}$ at 5556~\AA, 
dotted line with open circles: Hayes fluxes adjusted by 1.006
and wavelengths converted to vacuum to match STIS and the model. }
\end{figure}

\begin{figure}
\centering
\includegraphics[width=.4\textwidth,angle=90]{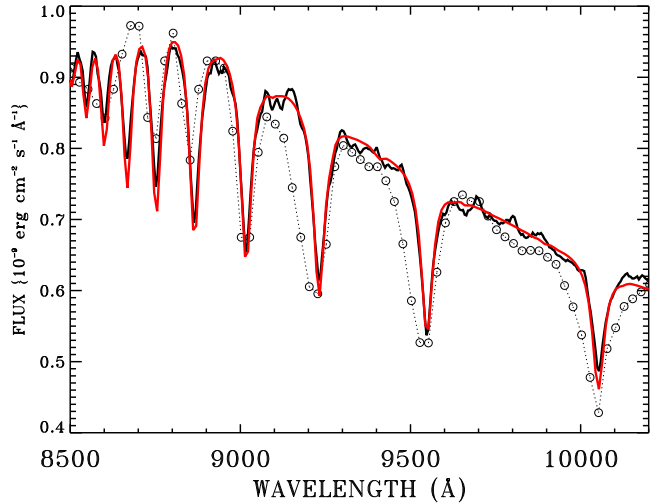}
\caption{As in Figure 3 for the region of the \ion{H}{1} Paschen lines, where the
Hayes (1985) spectrophotometry differs the most from STIS and the Kurucz model.
At these longest wavelengths of the STIS CCD data, the de-fringing technique
leaves some residuals at the $\sim$2\% level. The STIS data and $R=500$ model agree
within the uncertainty of the de-fringing.}
\end{figure}

\begin{figure}
\centering
\includegraphics[width=.5\textwidth]{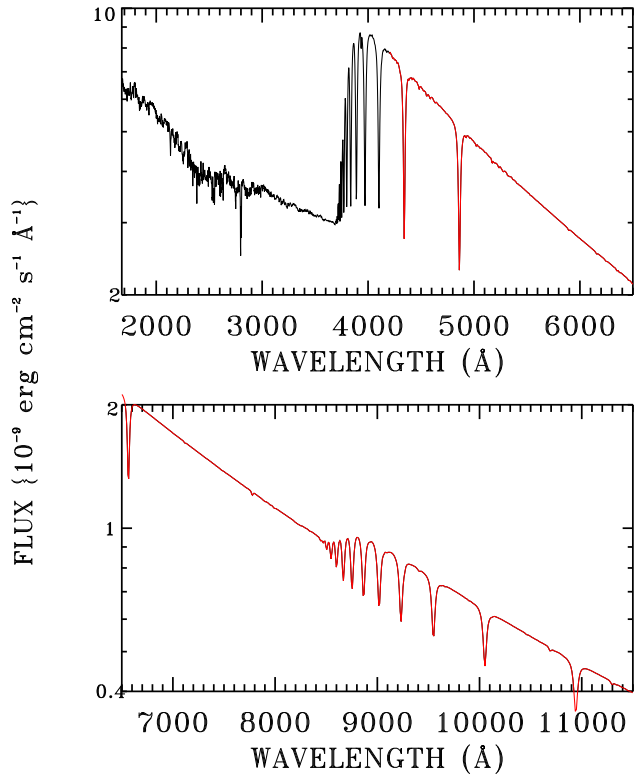}
\caption{Absolute flux distribution of Vega as measured by STIS below 4200~\AA\ 
and as determined by the Kurucz (2003) $R=500$ model at the longer wavelengths.
Below 3000~\AA\ the spectrum is dominated by metal line blanketing, while above
3000~\AA, only the Balmer and Paschen lines of hydrogen 
are deeper than $\sim$2\% at the
STIS resolution of $R=\sim500$. The top panel shows the peak of the flux distribution
down to $2\times10^{-9}$ erg cm$^{-2}$ s$^{-1}$ \AA$^{-1}$, while the 
lower panel covers the next factor of five lower fluxes.}
\end{figure}

\begin{figure}
\centering
\includegraphics[width=.6\textwidth,angle=90]{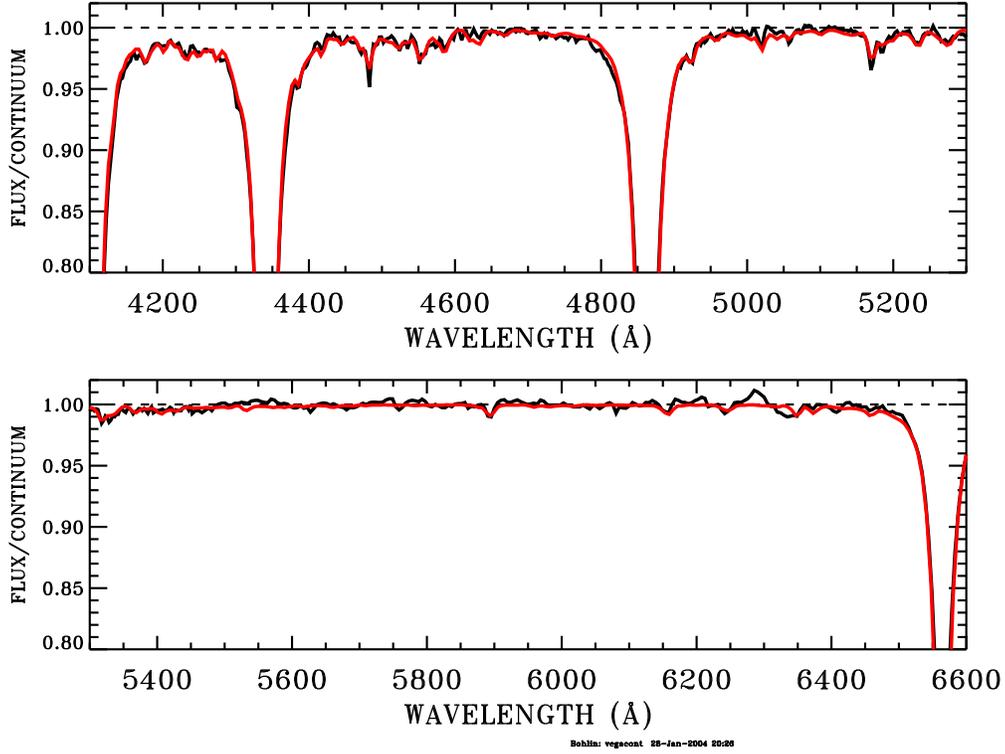}
\caption{Comparison of STIS (black) to the Kurucz (2003) model (red) that has
effective temperature 9550~K and $\log  g=3.95$ from H$\delta$ to H$\alpha$.
The observations and theory agree to $\sim$1\%.} \end{figure}

\begin{figure}
\centering
\includegraphics[width=.6\textwidth,angle=90]{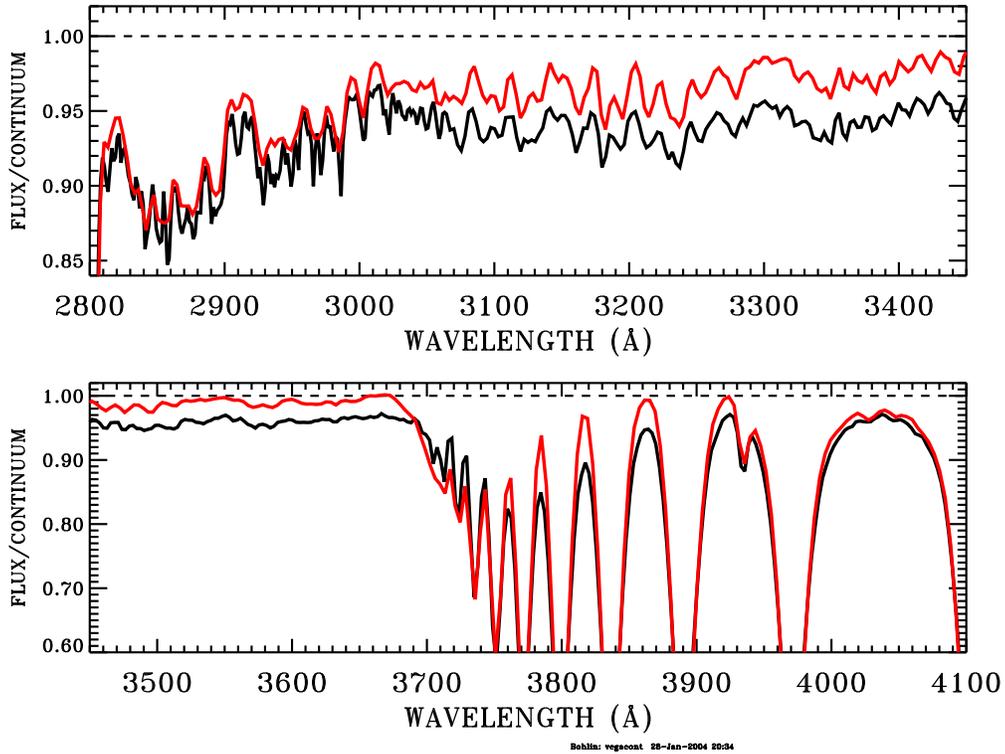}
\caption{Comparison as in Figure~6 for the Balmer continuum region. Systematic
differences of up to $\sim$3\% are prevalent. The upper panel has an expanded
vertical scale in comparison to the lower panel.}
\end{figure}

\begin{figure}
\centering
\includegraphics[width=.6\textwidth,angle=90]{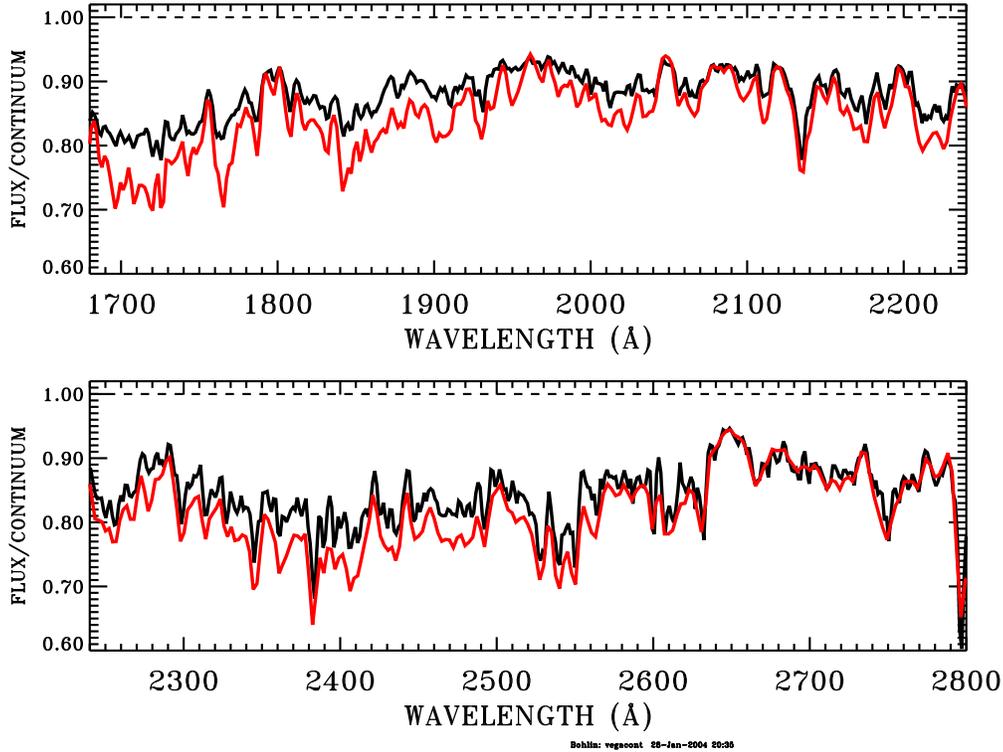}
\caption{Comparison in the UV below 2800~\AA, where there is an excellent
correlation in the wavelengths of the detailed spectral features. However, the
model often shows an excess of metal line blanketing by up to several percent.}
\end{figure}

\begin{deluxetable}{lllllcccc}
\tablewidth{0pt}
\tablecaption{Journal of Observations}
\tablehead{
\colhead{Root} &\colhead{Mode} &\colhead{Aper.}
&\colhead{Date}
&\colhead{Time} &\colhead{Propid} &\colhead{Exptime(s)}&\colhead{CR-split}
&\colhead{Postarg}\\
&&&&&&&&\colhead{(arcsec)}
}
\startdata
O8I105010  &G230LB  &52X2      &03-06-30 &11:00:47  &9664	 &0.9              &3     &0.5\\
O8I105020  &G230LB  &52X2      &03-06-30 &11:03:02  &9664	 &\llap{1}8.0   &3     &0.5\\
O8I105030  &G230LB  &52X2      &03-06-30 &11:08:37  &9664	 &0.9              &3     &0.0\\
O8I105040  &G230LB  &52X2E1  &03-06-30 &11:12:38  &9664	 &0.9              &3     &0.0\\
O8I105060  &G430L    &52X2       &03-06-30 &12:10:37  &9664	 &0.9              &3     &0.0\\
O8I105070  &G750L    &52X2       &03-06-30 &12:17:45  &9664	 &0.9              &3     &0.0\\
O8I106010  &G430L    &52X2       &03-08-23 &22:29:15  &9664	 &0.9              &3     &0.0\\
O8I106020  &G230LB  &52X2       &03-08-23 &22:36:23  &9664	 &0.9              &3     &0.0\\
O8I106030  &G230LB  &52X2       &03-08-23 &22:41:37  &9664	 &\llap{1}8.0   &3     &0.0\\
O8I106040  &G430L    &52X2        &03-08-23 &22:49:00  &9664	 &0.9               &3     &0.0\\
O8I106050  &G750L    &52X2        &03-08-23 &22:59:07  &9664	 &1.5               &5     &0.0\\
\enddata
\end{deluxetable}
\end{document}